\documentstyle[12pt]{article}

\begin{document}
\title{\textbf{Cosmic dynamo analogue and decay of magnetic fields in 3D Ricci flows}} \maketitle
{\sl \textbf{L.C. Garcia de Andrade}\newline
Departamento de F\'{\i}sica
Te\'orica-IF\newline
Universidade do Estado do Rio de Janeiro\\[-3mm]
Rua S\~ao Francisco Xavier, 524\\[-3mm]
Cep 20550-003, Maracan\~a, Rio de Janeiro, RJ, Brasil\\[-3mm]
Electronic mail address: garcia@dft.if.uerj.br\\[-3mm]
\vspace{0.01cm} \newline{\bf Abstract} \paragraph*{}Magnetic curvature effects, investigated by Barrow and Tsagas (BT) [Phys Rev D \textbf{77},(2008)],as a mechanism for magnetic field decay in open Friedmann universes (${\Lambda}<0$), are applied to dynamo geometric Ricci flows in 3D curved substrate in laboratory. By simple derivation, a covariant three-dimensional magnetic self-induced equation, presence of these curvature effects, indicates that de Sitter cosmological constant (${\Lambda}\ge{0}$), leads to enhancement in the fast kinematic dynamo action which adds to stretching of plasma flows. From the magnetic growth rate, the strong shear case, anti-de Sitter case (${\Lambda}<0$) BT magnetic decaying fields are possible while for weak shear, fast dynamos are possible. The self-induced equation in Ricci flows is similar to the equation derived by BT in $(3+1)$-spacetime continuum. Lyapunov-de Sitter metric is obtained from Ricci flow eigenvalue problem. In de Sitter analogue there is a decay rate of ${\gamma}\approx{-{\Lambda}}\approx{-10^{-35}s^{-2}}$ from corresponding cosmological constant ${\Lambda}$, showing that, even in the dynamo case, magnetic field growth is slower than de Sitter inflation, which strongly supports to BT result.

\newpage
\section{Introduction}
Recently Fields medalist Grisha Perelman \cite{1} has used the concept of Ricci flows , proposed by Hamilton in 1982 \cite{2}, to prove long standing unproved, Poincare conjecture on two and three-dimensional settings. Actually Perelman, argued that the Ricci flow could be immersed in a large spacetime structure, not necessarily relativistic. In these approach and Einstein and Jordan Brans-Dicke equations have been solved in this so-called, Ricci flow gravity. More recently, several attempts to generalize Ricci flows to gravity have been done mainly by Graf \cite{3} and Letelier \cite{4}. On the other hand, cosmology in the Laboratory (COSLAB) program developed mainly by Unruh, Visser, and Volovik \cite{5,6,7} has produced several papers on the analog models in general relativity (GR) and cosmology, which mimic in superfluid and other optical and hydrodynamics labs, the classical and quantum conditions in the universe. Yet in another front, the fast dynamos operating in solar physics and other astrophysical and cosmic settings, have shown the importance of dynamo theory in explaining, the magnetic field grow in the universe. In this paper, the cosmology analog of laboratory dynamos have been here proposed and derived from the Lyapunov metric exponents and Ricci flows equation in 3D. Dynamo stretching by Riemannian plasma curved substrates \cite{8,9} have led us , to this reasoning. The computation of the magnetic growth rate and covariant three-dimensional magnetic self-induced equation, shows that the presence of these curvature effects, indicates that de Sitter cosmological constant (${\Lambda}\ge{0}$), leads to enhancement in the fast dynamo action which adds to stretching of plasma flows. This result was proposed earlier by Barrow and Tsagas \cite{10} in the form of a slow, decaying magnetic field. The magnetic field growth rate is computed in terms of eigenvalues of the Ricci tensor in Einstein spaces \cite{11}. Besides reproducing the decaying magnetic result of BT in Ricci flows, fast dynamo action is also obtained obtained when the real part of the magnetic growth rate is positive. Note that stretching dynamic by plasma flows have also been obtained by M Nu\~nez \cite{8}. In this paper techniques of Einstein gravity, called Ricci rotation coefficients, are used to obtain the a Ricci flows dynamos. Note that a most important framework in this derivation is the proof that the Lyapunov metric exponents obtained by the eigenvalue problem of the Ricci flows leads naturally to a 3D cross section of the de Sitter spacetime, which support our conclusions. A detailed account of GR cosmological dynamos is contained in Widrow \cite{12}. The importance of laboratory analogues stems from the fact that there is no apparently stringy dynamo \cite{13}, and since even if there were, high energy physics at this level of energy is only in its infancy in CERN experiments. The resulting metric is called de Sitter-Lyapunov metric. Also recently Fedichev and Fischer \cite{14} investigated a quasi-particle cosmic analogue, using also de Sitter equivalents to trapp particles. Actually BT attempt is not the first one to relate the magnetisation coupling to curvature invariants, a similar attempt has been done by Bassett et al \cite{15}, which has related preheating phases of the universe to magnetic dynamo. In the de Sitter case a simple computation shows that, the growth rate of the magnetic field is slower than the expansion of the model, and shear is exactly the agent that slows down the dynamo growth. Thus one can say that cosmic fast dynamo is slower down by the shear of the cosmic Ricci flow. The paper is organized as follows: Section II presents the Ricci flow as a de Sitter-Lyapunov metric and section III, derivation of the self-induction equation in Ricci flows is given and the cosmic dynamo analogue is presented. Section IV presents future prospects and conclusions.
\newpage
\section{De Sitter-Lyapunov analogue metric in Ricci flows}
In this section, though more mathematical than the rest of paper , is fundamental to understand why the relation between de Sitter space 3D section appears so naturally in the context of the Lyapunov metric exponents which leads to chaotic dynamos, which are so interesting from the cosmological point of view. The Riemann metric given by the Ricci flow \cite{1,2}, is given in mathematical terms by\newline
\textbf{Definition II.1}:
\begin{equation}
\frac{{\partial}\textbf{g}}{{\partial}t}=-2\textbf{Ric}
\label{1}
\end{equation}
where here, $\textbf{g}$ is the Riemann metric, over manifold $\cal{M}$, and the parameter t in the Riemann metric $\textbf{g}(t)$, is given in the interval $t\in{[a,b]}$ in the field of real numbers $\textbf{R}$. On a local chart $\cal{U}$ in $\cal{M}$, the expression (\ref{1}) can be expressed as \cite{18}
\begin{equation}
\frac{{\partial}{g_{ij}}}{{\partial}t}=-2{R_{ij}}
\label{2}
\end{equation}
where $\textbf{Ric}$, is the Ricci tensor, whose components $R_{ij}$. From this expression, one defines the eigenvalue problem as
\begin{equation}
R_{ij}{\chi}^{j}={\lambda}{\chi}_{i}\label{3}
\end{equation}
where $(i,j=1,2,3)$. Substitution of the Ricci flow equation (\ref{2}) into this eigenvalue expression and cancelling the eigendirection ${\chi}^{i}$ on both sides of the equation yields
\begin{equation}
\frac{{\partial}g_{ij}}{{\partial}t}=-2\lambda{g_{ij}}
\label{4}
\end{equation}
Solution of this equation yields the Lyapunov expression for the metric
\begin{equation}
g_{ij}=exp{[-2\lambda{t}]}{\delta}_{ij}
\label{5}
\end{equation}
where ${\delta}_{ij}$ is the Kroenecker delta. Note that in principle if ${\lambda}\le{0}$ the metric grows without bounds, and in case it is negative it is bounded as $t\rightarrow{\infty}$. Recently Thiffeault has used a similar Lyapunov exponents expression in Riemannian manifolds to investigate chaotic flows, without attention to dynamos or Ricci flow. Thus one has proven the following lemma:\newline
\textbf{Lemma II.1}:\newline
If ${\lambda}_{i}$ is an eigenvalue spectra of the $\textbf{Ric}$ tensor, the finite-time Lyapunov exponents spectra is given by
\begin{equation}
{\lambda}_{i}=-{\gamma}_{i}\le{0}\label{6}
\end{equation}
In the next section I shall use this argument to work with the de Sitter metric
\begin{equation}
ds^{2}=-dt^{2}+e^{{\Lambda}t}(dx^{2}+dy^{2}+dz^{2})\label{7}
\end{equation}
in the de Sitter-Lyapunov analogue 3D spacetime
\begin{equation}
ds^{2}=e^{{\Lambda}t}(dx^{2}+dy^{2}+dz^{2})\label{8}
\end{equation}
which shows that the de Sitter-Lyapunov metric can be obtained from de Sitter metric by simply consider a constant slice $t=constant$ of de Sitter spacetime. Nevertheless, it is important to recall that for practical uses small variations of the parameter t would be allowed, otherwise the metric should be Ricci flat since de Sitter-Lyapunov metric should be flat, if $t-constant$.

\section{Ricci dynamo flows as a 3D cosmic analogue}
Now let us consider, Lyapunov eigenvalues, shall play an important role in the determination of the bounds of magnetic energy as a global dynamo action bound.
Let us now consider the magnetic kinematic dynamo, most commonly known as an equation, with non-zero plasma resistivity ${\eta}$
\begin{equation}
\frac{d\textbf{B}}{dt}=\textbf{B}.{\nabla}\textbf{v}+{\eta}{\Delta}\textbf{B}\label{9}
\end{equation}
where $\textbf{B}$ is the magnetic field vector, and $\textbf{v}$ is the flow velocity, where ${\Delta}:={\nabla}^{2}$ is the Laplacian operator. Here we also assume that the covariant flow derivative is given by
\begin{equation}
\frac{d\textbf{}}{dt}=\textbf{v}.{\nabla}+\frac{{\partial}}{{\partial}t}\label{10}
\end{equation}

Here we assume that the magnetic self-induction equation (\ref{9}) is non-relativistic since we are not in the true GR but in laboratory analogue cosmology. A long but straightforward computation, yield the diffusion term as
\begin{equation}
{\Delta}\textbf{B}= \frac{1}{\sqrt{g}}{\partial}_{i}[\sqrt{g}g^{ij}{\partial}_{j}\textbf{B}]\label{11}
\end{equation}
which expanded using the frame ${\textbf{e}}_{i}$ where $(i,j=1,2,3)$ and
\begin{equation}
\textbf{B}=B^{i}\textbf{e}_{i}\label{12}
\end{equation}
yields
\begin{equation}
{\Delta}\textbf{B}= [{g}^{ij}{\partial}_{i}{\partial}_{j}B^{p}+B^{k}[{\partial}_{i}{{\gamma}^{p}}_{jk}g^{ij}+
{{\gamma}^{l}}_{jk}{{\gamma}^{p}}_{il}g^{ij}]+[{{\gamma}^{p}}_{jk}g^{ij}{\partial}_{i}B^{k}]]
\textbf{e}_{p}\label{13}
\end{equation}
Here ${{\gamma}^{l}}_{jk}$ is the Ricci rotation coefficients (RRCs) analogous to the Riemann-Christoffel symbols. The RRCs is defined by
\begin{equation}
{{\partial}_{k}}\textbf{e}_{i}={{\gamma}_{ki}}^{j}\textbf{e}_{j}\label{14}
\end{equation}
The Christoffel symbols
\begin{equation}
{{\Gamma}^{i}}_{jk}=g^{il}[{{\partial}_{j}}g_{kl}+{{\partial}_{k}}g_{jl}-
{\partial}_{l}g_{jk}]\label{15}
\end{equation}
do not appear in the computations, since we have assumed, that the trace of the Christoffel symbols vanish. To complete the derivation of the self-induction equation it remains to obtain the diffusion free part of the self-induced equation above , which in general curvilinear coordinates ${x^{i}}\in{\cal{U}}_{i}$, of the sub-chart $U_{i}$ of the manifold, in the rotating frame reference of the flow $\textbf{e}_{i}$, reads
\begin{equation}
\frac{d\textbf{B}}{dt}=({\textbf{B}}.{\nabla})\textbf{v}\label{16}
\end{equation}
Before this derivation, let us now introduce the Ricci tensor into play, by considering the following trick
\begin{equation}
\frac{d}{dt}[g^{il}g_{lk}]=\frac{d}{dt}[{{\delta}^{i}}_{k}]=0\label{17}
\end{equation}
which can be applied to the expression
\begin{equation}
\frac{d\textbf{B}}{dt}= \frac{d}{dt}(g^{ik}B_{k}\textbf{e}_{i})\label{18}
\end{equation}
to obtain
\begin{equation}
\frac{d}{dt}{(g^{ik}B_{k})}= \frac{d}{dt}(g^{ik})B_{k}+g^{ik}\frac{d}{dt}{B_{k}}\label{19}
\end{equation}
Now by making use of the Ricci flow equation above into this last expression, yields
\begin{equation}
\frac{d}{dt}{g^{ik}B_{i}}= -2R^{ik}B_{k}+g^{ik}\frac{d}{dt}{B_{k}}\label{20}
\end{equation}
Note that for de Sitter spacetime the solenoidal magnetic field is also satisfied and no magnetic monopole is assumed in this phase, thus 
\begin{equation}
{\nabla}.\textbf{B}=0 \label{21}
\end{equation}
From the evolution of the reference frame
\begin{equation}
\frac{d\textbf{e}_{i}}{dt}={{\omega}_{i}}^{j}\textbf{e}_{j}\label{22}
\end{equation}
and the Ricci rotation coefficient, one obtains the magnetic curvature effect in dynamo theory, through the self-induction equation in Ricci flow as
\begin{equation}
[{\gamma}+2{\Lambda}+{\omega}]B_{i}=B^{p}[v^{l}{{\gamma}^{i}}_{pl}+v_{j}{\partial}_{p}g^{ij}+
g^{ij}[{\sigma}_{pj}+{\Omega}_{pj}-\frac{1}{3}{\theta}{g}_{pj}]]\label{23}
\end{equation}
where we have used the following expressions
\begin{equation}
{{\partial}_{k}}\textbf{e}_{i}={{\Gamma}_{ki}}^{j}\textbf{e}_{j}\label{24}
\end{equation}
where ${\omega}_{ij}$ and ${{\Gamma}_{ki}}^{j}$ are respectively the vorticity, and the gradient of the Ricci flow $v^{l}$ is decomposed into its invariant format of vorticity ${\Omega}_{lp}$, shear ${\sigma}_{kl}$ tensors and expansion ${\theta}$
as
\begin{equation}
{\partial}_{p}v_{l}={\Omega}_{pl}+{\sigma}_{pl}-\frac{1}{3}{\theta}g_{lk}
\label{25}
\end{equation}
These equations were also simplified by the assumption that the flow has a rigid rotation, or that the vorticity of the flow coincides with the vorticity of the frame or
\begin{equation}
{\Omega}_{pl}={\omega}_{pl}
\label{26}
\end{equation}
This assumption is cosmologically reasonable, since the in inflationary cosmological models the vorticity and shear are smaller than the expansion, which represents the stretching in the language of dynamo theory. The fact that the expansion is the trace of the gradient strain ${\theta}=Tr[{\nabla}\textbf{v}]$, shows that the Ricci dynamo flow is compressible, which are more pathological dynamo flows that the ones that are compressible, or solenoidal
\begin{equation}
{\nabla}.\textbf{v}=0
\label{27}
\end{equation}
To deduce expression (\ref{23}) we still have used the eigenvalue expressions
\begin{equation}
{\sigma}_{ki}B^{i}={\sigma}B_{k}
\label{28}
\end{equation}
\begin{equation}
{\Omega}_{ki}B^{i}={\Omega}B_{k}
\label{29}
\end{equation}
\begin{equation}
{\Theta}_{ki}B^{i}=-\frac{1}{3}{\theta}B_{k}
\label{30}
\end{equation}
for the kinematical cosmological Ehlers-Sachs quantities. Besides one also uses the fact that the Ricci flows obey the Einstein manifold 3D condition
\begin{equation}
R_{lp}={\Lambda}g_{lp}
\label{31}
\end{equation}
With all those simplifications the dynamo equation (\ref{23}) allows us to compute the dynamo growth rate as
\begin{equation}
{\gamma}=[2{\Lambda}-{\sigma}+\frac{1}{3}{\theta}]+\frac{B^{p}B_{i}v_{l}}{B^{2}}[{{{\gamma}^{i}}_{p}}^{l}
+{\partial}_{p}g^{il}]
\label{32}
\end{equation}
From this expression, one immeadiatly notices that the stretching term contributes to enhance dynamo action, while the positive ${\gamma}$ de Sitter cosmological constant, also enhances the fast dynamo action while the anti-de Sitter or open Friedmann universe induces the decay of magnetic field or the BT result \cite{10}. The last expression comes from an expression with the Ricci tensor similar to the BT wave equation, which we repeat here for readers convenience
\begin{equation}
\frac{d^{2}B_{i}}{d{\tau}^{2}}-D^{2}{B_{i}}=-5{\Lambda}\frac{d}{d{\tau}}
{B_{i}}-4{\Lambda}^{2}B_{i}+\frac{1}{3}({\rho}+3p)B_{i}-R_{ij}B^{j}
\label{33}
\end{equation}
In their notation $B_{i}$ is the magnetic vector field in the comoving frame, and $D^{2}=h_{ij}{\nabla}^{i}{\nabla}^{j}$ is the 3D Laplacian, where $h_{ij}=g_{ij}+v_{i}v_{j}$ is the projection metric orthogonal to $v^{l}$. 
Note that in de Sitter spacetime, the growth rate of the dynamo action yields
This expression allows us to obtain the growth rate ${\gamma}$ defined as
\begin{equation}
{\gamma}=[2{\Lambda}-{\sigma}+\frac{1}{3}{\theta}]
\label{34}
\end{equation}
since the terms ${{{\gamma}^{i}}_{p}}^{l}$ and ${\partial}_{p}g^{ij}$ vanish for the de Sitter metric components $g^{ij}=e^{-{\Lambda}t}{\delta}^{ij}$. Let us now compute the growth rate of the cosmic Ricci dynamo flow in the case of the 3D section of the Friedmann-Robertson-Walker (FRW) universe
\begin{equation}
dl^{2}=\frac{dr^{2}}{(1-\frac{{\Lambda}r^{2}}{2})}+r^{2}d{\Omega}^{2}
\label{35}
\end{equation}
where $d{\Omega}^{2}=(d{\theta}^{2}+sin^{2}{\theta}d{\phi}^{2})$ is the solid angle. Computation of the Ricci rotation coefficients in the limit of $r\rightarrow{0}$, yields
\begin{equation}
{\gamma}=[2{\Lambda}-{\sigma}+\frac{1}{3}{\theta}]-2(1+\frac{{\Lambda}r^{2}}{2})v^{r}
\label{36}
\end{equation}
Note that not only shear, slows down the dynamo action, but in FRW universe, since the cosmological constant is very small, the the last term does not enhance dynamo action and magnetic fields decay in this universe model.
Taking into account the magnetic energy ${\epsilon}$ as
\begin{equation}
{\epsilon}=\int{B^{2}dV}
\label{37}
\end{equation}
which expressed in terms of the 3D Riemann metric components reads
\begin{equation}
{\epsilon}=\int{B^{i}g_{ij}B^{j}dV}
\label{38}
\end{equation}
Since, by definition fast dynamo action corresponds to the growth of magnetic energy in time as $\frac{{\partial}{\epsilon}}{{\partial}t}\ge{0}$ , this amount has to be computed by performing the partial time derivative of the expression (\ref{25}). Actually the equal sign in the last condition represents the lower limit of marginal dynamos, where the magnetic energy integral remains constant. This computation yields
\begin{equation}
\frac{{\partial}{\epsilon}}{{\partial}t}=\frac{{\partial}[\int{B^{i}g_{ij}B^{j}dV}]}{{\partial}t}
\label{39}
\end{equation}
Expansion of the RHS of this expression shows clearly now where the Ricci flow eigenvalue effect is going to appear. A simple computation, shows that the energy integral confirms the dynamo action. Throughout the paper the diffusion term was not explicitly computed since because we use the limit of diffusion free to check for the presence of slow dynamos, which seems not exist globally in the universe.
Note that the in the de Sitter case the magnetic can be written as
\begin{equation}
B^{i}=B^{0}e^{[2{\Lambda}-{\sigma}+\frac{1}{3}{\theta}]t}
\label{40}
\end{equation}
which shows that the shear eigenvalue ${\sigma}$, slow down dynamo action, while the cosmological constant in de Sitter space enhances it. Anti-de Sitter effective \cite{3} spacetime, of course contributes to slow down magnetic field as negative exponents contributes to the decay of magnetic field in the effective universe.
\section{Conclusions}
 By making use of mathematical tools from Riemannian geometry, so popularised in Einstein general relativity, called Ricci Rotation Coefficients, one obtains fast dynamo action in stretching magnetic field lines endowed with shear in de Sitter-Lyapunov analogue spacetime metric. Besides the fast dynamo action for de Sitter or closed (3+1)-spacetime Ricci flows, where the cosmological constant ${\Lambda}>0$, which is a new result, one is able to reproduce the BT magnetic field decay in $(3+1)$ real spacetime of GR cosmology. This seems to shed some light on the implications of Ricci flow to more generalised settings and take it out from the pure mathematical applications. Note that innumeral applications of Ricci flows to physics have been considered so far, but this is the first time, to our knowledge, that it is applied to cosmological analogues. An interesting panorama of the applications of Ricci flows manifolds in Physics maybe found in the paper by Woolgar \cite{16}. Ricci flows are mainly applied in solitons and since this is an important subject to cosmology \cite{17} we may address the relation between our cosmic analogues Ricci flows to solitons. This may appear elsewhere. Another interesting perspective for Ricci dynamo flows is to investigate them in the light of recently proposed COBE fractal geometry analysis, of Caruso and Oguri \cite{18}. After I finish this paper, it came to my knowledge that Marklund and Clarkson \cite{19} have presented a general GR covariant formalismo for dynamo magnetohydrodynamics equation, where where however, no Ricci flows are present and only gravitational waves application in diffusive plasma are given. Vortex dynamos in analogue models can be also treated elsewhere, based on a non-Riemannian vortex acoustics model presented earlier by the author \cite{20}. Recent magnetic flux tubes in Riemannian manifolds \cite{21} may also be addressed in the Ricci flow dynamo context.\section{Acknowledgements}
 Several discussions with J L Thiffeault are highly appreciated. I also thank financial  supports from UERJ and CNPq.
 

\newpage

\begin{thebibliography}{21}
  \bibitem{1} G Perelmann, The entropy formula for the Ricci flow and its geometric applications , Los Alamos arxives math-DG /0211159.
  \bibitem{2} R S Hamilton, J Diff Geom, \textbf{17},255 (1982). B Chow and D Knopf, Ricci Flows:An introduction, (2004) AMS, New York.
  \bibitem{3} W. Graf, \textbf{Ricci Flow Gravity},gr-qc/0602054v3. 
  \bibitem{4} P S Letelier, Int J Theor Phys \textbf{A 43}, 4281 (1991).
  \bibitem{5} G Volovik, Universe in a Helium Droplet (2003) Oxford University Press.
  \bibitem{6} W Unruh, Phys Rev Lett \textbf{46},1351(1981).
  \bibitem{7} U Fischer and M Visser, Phys Rev Lett (2001).
  \bibitem{8} J L Thiffeault, Differential constraints in Chaotic Flows on Curved Manifolds, (2002) arXiv:nlin/0209042.v1.
  \bibitem{9} M Nu\~nez, J Phys \textbf{A},8903 (2003).
  \bibitem{10} J D Barrow and C Tsagas, Phys Rev D \textbf{77},107302 (2008). 
  \bibitem{11} A Besse, Einstein manifolds, (2007) Springer-Verlag.
  \bibitem{12} L Widrow, Rev Mod Phys \textbf{74} (2002) 775.
  \bibitem{13} A Vilenkin and A Shellard, \textbf{Topological defects in spacetime} (2000) Cambridge University Press.
  \bibitem{14} Yu Fedichev and U Fischer, Phys Rev A \textbf{69}, 0303063 (2004).
  \bibitem{15} B Bassett,G Pollifrone, S Tsujikawa and F Viniegra, \textbf{Pre-heating: Cosmic dynamo?}.arxiv:astro-ph/0010628v3.
  \bibitem{16} Yu Fedichev and U Fischer, Phys Rev A \textbf{69}, 0303063 (2004).
  \bibitem{17} E. Woolgar, Some applications of the Ricci flows in Physics, Los Alamos arxiv: 0708.2144.
  \bibitem{18} F Caruso and V Oguri, Ap J, \textbf{694},151 (2009).
  \bibitem{19} M Marklund and C Clarkson, Mont Not Roy Astron Soc (2004).
  \bibitem{20} L C Garcia de Andrade, Phys Rev D \textbf{70}.
  \bibitem{21} L C Garcia de Andrade, The role of stretching and
  curvature in fast dynamo plasmas in Riemannian space, Phys Plasmas \textbf{15} (2008) in press.
  \end{thebibliography}
  \end{document}